\documentclass[12pt]{iopart} 

\begin{document}

\title{Acceleration of the universe in the Einstein frame 
of a metric-affine $f(R)$ gravity} 

\author{Nikodem J. Pop\l awski}

\address{Department of Physics, Indiana University, 
727 East Third Street, Swain Hall West 117, Bloomington, 
IN 47405-7105, USA}
\ead{nipoplaw@indiana.edu}

\begin{abstract}
We show that inflation and current cosmic acceleration can be generated by a 
metric-affine $f(R)$ gravity formulated in the Einstein conformal frame, 
if the gravitational Lagrangian $L(R)$ contains both positive and 
negative powers of the curvature scalar $R$. 
In this frame, we give the equations for the 
expansion of the homogeneous and isotropic matter-dominated universe 
in the case $L(R)=R+\frac{R^3}{\beta^2}-\frac{\alpha^2}{3R}$, where 
$\alpha$ and $\beta$ are constants.
We also show that gravitational effects of matter in such a universe 
at very late stages of its expansion are
weakened by a factor that tends to $3/4$, 
and the energy density of matter $\epsilon$ scales the same way as in the $\Lambda CDM$ model only when $\kappa\epsilon\ll\alpha$. 
\end{abstract}

\pacs{04.50.+h, 98.80.-k}

\maketitle

\section{Introduction} 
Recent observations of type Ia supernovae~\cite{SN} and the cosmic 
microwave background radiation~\cite{CMB1,CMB2} indicate that the 
universe is undergoing a phase of accelerated expansion. 
It is also believed that shortly after the big bang, the universe passed
through a phase of extremely rapid expansion (inflation)~\cite{infl}.
Such an idea resolves several problems in big bang cosmology,
e.g., the flatness problem and the horizon problem.
The most accepted explanation of current cosmic acceleration is that the 
universe is dominated by dark energy~\cite{DE}, whereas inflation
is thought to be driven by scalar fields with an appropriate potential. 

However, it is also possible to modify Einstein's general relativity 
to obtain gravitational field equations that allow both inflation and present
cosmic acceleration. 
A particular class of alternative theories of gravity that has recently 
attracted a lot of interest is that of the $f(R)$ gravity 
models, in which the gravitational Lagrangian is a function of 
the curvature scalar $R$~\cite{Bar}. 
It has been shown that current acceleration may originate from the addition of 
a term $R^{-1}$ (or other negative powers of $R$) to the Einstein-Hilbert
Lagrangian $R$, whereas terms with positive powers of $R$ may be the source 
of inflation~\cite{fR1,fR2}.
As in general relativity, these models obtain the field equations by 
varying the total action for both the field and matter.

There are two different approaches for how to vary the action in these
models: metric and metric-affine.
The first one is the usual Einstein-Hilbert variational principle, 
according to which the action is varied with respect to the metric 
tensor 
$g_{\mu\nu}$, and the affine connection is given by the Christoffel symbols 
(the Levi-Civita connection)~\cite{LL}. 
The other one is the Palatini variational principle 
(originally formulated by Einstein), according to which the metric and 
the connection are considered as geometrically independent quantities, 
and the action must be varied with respect to both of 
them~\cite{Pal1}. 
Both the metric and the metric-affine approaches give the same result 
only if we use the standard Einstein-Hilbert action, since  
variation with respect 
to the connection gives the usual expression for the Christoffel 
symbols.

Of the two approaches, the metric-affine formalism seems more 
general since it requires 
one less constraint than the metric approach
(no {\it a priori} relation between the metric and the connection).
Moreover, the field equations in this formalism are second-order 
differential equations (and the Cauchy problem is similar to that
in general relativity), whereas in metric theories they are 
fourth-order~\cite{eq1}. 
Another remarkable property of the metric-affine approach is that the 
field equations in vacuum reduce to the standard Einstein equations of 
general relativity with a cosmological constant~\cite{univ},
whereas vacuum metric theories allow both accelerated and
decelerated phases of the universe expansion~\cite{nom}. 
This approach is also free of instabilities that appear in the metric 
formulation of $1/R$ gravity~\cite{inst1}, although such instabilities
can be suppressed by adding to the Lagrangian terms with positive
powers of $R$~\cite{inst2}, or by quantum effects~\cite{inst3}.
Furthermore, there is a debate on the compatibility of $f(R)$ gravities 
with solar system observations~\cite{sol}, and on the Newtonian 
limit of these theories~\cite{lim1}. 
There is theoretical evidence suggesting that metric-affine models 
with the inverse power of the curvature scalar have a good Newtonian 
limit~\cite{lim2}, and that metric models pass the solar system 
tests~\cite{inst2,lim3}.

One can show that any of these theories of gravitation 
is conformally equivalent to the Einstein theory
of the gravitational field interacting with additional matter 
fields~\cite{eq2}. 
However, a GR-like formulation can be obtained without any 
redefinition of the metric, by isolating the spin-0 degree of freedom 
due to the occurrence of nonlinear second-order terms in the Lagrangian, 
and encoding 
it into an auxiliary scalar field $\phi$ by means of a Legendre 
transformation. Such a transformation in classical mechanics replaces 
the Lagrangian of a mechanical system with the Helmholtz 
Lagrangian~\cite{eq3}. 

The set of variables $(g_{\mu\nu},\,\phi)$ is commonly called the 
{\it Jordan conformal frame}, although it refers to a choice of 
dynamical 
variables rather than to a choice of a frame of reference. 
In the Jordan frame of metric $f(R)$ theories, the self-gravitating scalar 
field $\phi$ violates the stability of vacuum and the positivity of 
energy~\cite{eq1,eq3}. 
These unphysical properties can be eliminated by a certain conformal 
transformation of the metric: $g_{\mu\nu}\rightarrow 
h_{\mu\nu}=f'(R)g_{\mu\nu}$. 
The new set $(h_{\mu\nu},\,\phi)$ is called the {\it Einstein conformal 
frame}. 
Although both frames are equivalent mathematically, they are {\it not}  
equivalent physically~\cite{EJ2}, and 
the interpretation of cosmological observations can drastically 
change depending on the adopted frame~\cite{EJ1}.
The physically measured metric is determined from the coupling
to matter, and the principle of equivalence can provide
an operational definition of the metric tensor~\cite{Br}.
Furthermore, in the Einstein frame, the principle of equivalence is
violated~\cite{EJ2} and the available tests of this principle can serve as 
the constraints on nonlinear gravities.

In the metric-affine formalism, the auxiliary field $\phi$ has no
kinetic term and does not violate the positivity of energy.
Therefore, which frame is physical is a matter of choice, although
this question should be ultimately answered by experiment or observation.
Remarkably, it is the Einstein frame in which the connection 
is metric-compatible.
Therefore, in this work we treat $h_{\mu\nu}$ as the physical 
metric tensor~\cite{eq1,eq3,EJ2}. 

The authors of~\cite{fR1,fR2} applied the metric 
variational formalism to $f(R)$ theories of gravitation.  
It has been shown that positive and 
negative powers of the curvature scalar in the gravitational Lagrangian
can cause inflation and current acceleration also in the metric-affine
formalism~\cite{Pal2,Pal8,Pal5,Pal6}, although 
the compatibility of Palatini $f(R)$ 
models with experiment is being debated~\cite{Pal3,Pal4},
and these models may face the problem of stability of matter
perturbations~\cite{Koi2}.
The simplest Lagrangian in a metric-affine theory, which drives both phases 
of acceleration, 
has the form $L(R)=R+\frac{R^3}{\beta^2}-\frac{\alpha^2}{3R}$~\cite{Pal7}. 
Here, $\alpha$ and $\beta$ are constants, and a cubic term was chosen
because a quadratic term $R^2$ cannot lead to gravity driven 
inflation in the Palatini formalism~\cite{MW}. 
We emphasize that, in both formalisms, the above authors studied 
cosmology in the {\it Jordan frame}.   

In this paper, we show that inflation and present cosmic acceleration can be 
generated by a metric-affine $f(R)$ gravity formulated in the
{\it Einstein frame}, if the Lagrangian contains both positive and 
negative powers of the curvature scalar. 
In this frame, we give explicit formulae for the expansion of the 
homogeneous and isotropic matter-dominated universe, using the Lagrangian
of~\cite{Pal7}.
We also show that gravitational effects of matter in such a universe 
at late stages of its expansion are
weakened, and the energy density of matter differs in scaling 
from that in the $\Lambda CDM$ model.

In section~2, we introduce the metric-affine formalism for an $f(R)$ 
gravity in the Einstein frame. 
In section~3, we apply the gravitational 
field equations to a homogeneous and isotropic universe, and study them 
for the above Lagrangian. 
The results are summarized in section~4.

\section{Metric-affine formalism in the Einstein conformal frame}
In this section we review the metric-affine variational approach to a 
gravitational theory~\cite{OK}. 
The equations of the field are obtained from the Palatini variational 
principle, according to which both the metric tensor $g_{\mu\nu}$ and 
the affine connection $\Gamma^{\,\,\rho}_{\mu\,\nu}$ are regarded as 
independent variables. 
The action for an $f(R)$ gravity in the original (Jordan) frame is 
given by
\begin{equation}
S_J=-\frac{1}{2\kappa c}\int d^4 
x\bigl[\sqrt{-\tilde{g}}L(\tilde{R})\bigr] + 
S_m(\tilde{g}_{\mu\nu},\psi).
\label{action1}
\end{equation}
Here, $\sqrt{-\tilde{g}}L(\tilde{R})$ is a Lagrangian density that depends 
on the curvature scalar $\tilde{R}$, $S_m$ is the action for matter 
represented 
symbolically by $\psi$, and $\kappa=\frac{8\pi G}{c^4}$. 
Tildes indicate quantities calculated in the Jordan frame.
The curvature scalar is obtained by contracting the Ricci tensor 
$R_{\mu\nu}$ with the metric tensor,
\begin{equation}
\tilde{R}=R_{\mu\nu}(\Gamma)\tilde{g}^{\mu\nu},
\label{scalar}
\end{equation}
and the Ricci tensor depends on the symmetric connection 
$\Gamma^{\,\,\rho}_{\mu\,\nu}$,
\begin{equation}
R_{\mu\nu}(\Gamma)=\partial_{\rho}\Gamma^{\,\,\rho}_{\nu\,\mu}-\partial_{\nu}\Gamma^{\,\,\rho}_{\rho\,\mu}+\Gamma^{\,\,\sigma}_{\nu\,\mu}\Gamma^{\,\,\rho}_{\rho\,\sigma}-\Gamma^{\,\,\sigma}_{\rho\,\mu}\Gamma^{\,\,\rho}_{\nu\,\sigma}.
\label{Ricci}
\end{equation}

Variation of the action with respect to $g_{\mu\nu}$ 
yields the field equations,
\begin{equation}
L'(\tilde{R})R_{\mu\nu}-\frac{1}{2}L(\tilde{R})\tilde{g}_{\mu\nu}=\kappa 
T_{\mu\nu},
\label{field}
\end{equation} 
where the dynamical energy-momentum tensor of matter is generated 
by the Jordan metric tensor,
\begin{equation}
\delta S_m=\frac{1}{2c}\int d^4 x\sqrt{-\tilde{g}}\,T_{\mu\nu}\delta\tilde{g}^{\mu\nu},
\label{EMT1}
\end{equation} 
and the prime denotes the derivative of a function with respect to
its argument.
If we assume that $S_m$ is independent of 
$\Gamma^{\,\,\rho}_{\mu\,\nu}$, then variation of the action with 
respect to the connection leads to
\begin{equation}
\nabla_{\rho}\Bigl(L'(\tilde{R})\tilde{g}^{\mu\nu}\sqrt{-\tilde{g}}\,\Bigr)=0,
\label{connect}
\end{equation} 
from which it follows that the affine connection coefficients are the Christoffel symbols, 
\begin{equation}
\Gamma^{\,\,\rho}_{\mu\,\nu}=\{^{\,\,\rho}_{\mu\,\nu}\}_g=\frac{1}{2}g^{\rho\lambda}(\partial_{\mu}g_{\nu\lambda}+\partial_{\nu}g_{\mu\lambda}-\partial_{\lambda}g_{\mu\nu}),
\label{Chr}
\end{equation}
with respect to the conformally transformed metric 
\begin{equation}
g_{\mu\nu}=L'(\tilde{R})\tilde{g}_{\mu\nu}.
\label{conf}
\end{equation}
The metric $g_{\mu\nu}$ (denoted by $h_{\mu\nu}$ in section~1) defines the Einstein frame with a geodesic structure (metric-compatible connection).

One can show that the action~(\ref{action1}) is dynamically equivalent 
to the following Helmholtz action~\cite{eq1,eq3},
\begin{equation}
S_H=-\frac{1}{2\kappa c}\int d^4 x\sqrt{-\tilde{g}}\bigl[L(\phi(p))+p(\tilde{R}-\phi(p))\bigr]+S_m(\tilde{g}_{\mu\nu},\psi),
\label{action2}
\end{equation}
where $p$ is a new scalar variable.
The function $\phi(p)$ is determined by
\begin{equation}
\frac{\partial 
L(\tilde{R})}{\partial\tilde{R}}\bigg{\vert}_{\tilde{R}=\phi(p)}=p.
\label{phi}
\end{equation}
From equations~(\ref{conf}) and~(\ref{phi}) it follows that
\begin{equation}
\phi=RL'(\phi).
\label{resc}
\end{equation}

In the Einstein frame, the action~(\ref{action2}) becomes the standard 
Einstein-Hilbert action of general relativity with additional non-kinetic 
scalar field terms:
\begin{equation}
S_E=-\frac{1}{2\kappa c}\int d^4 x\sqrt{-g}\bigl[R-\frac{\phi(p)}{p}+\frac{L(\phi(p))}{p^2}\bigr]+S_m(p^{-1}g_{\mu\nu},\psi).
\label{action3}
\end{equation}
Here, $R$ is the curvature scalar of the metric $g_{\mu\nu}$.
Choosing $\phi$ as the variable leads to
\begin{equation}
S_E=-\frac{1}{2\kappa c}\int d^4 x\sqrt{-g}\bigl[R-2V(\phi)\bigr]+S_m([L'(\phi)]^{-1}g_{\mu\nu},\psi),
\label{action4}
\end{equation}
where $V(\phi)$ is the effective potential,
\begin{equation}
V(\phi)=\frac{\phi L'(\phi)-L(\phi)}{2[L'(\phi)]^2}.
\label{pot}
\end{equation}

Variation of the action~(\ref{action4}) with respect to 
$g_{\mu\nu}$ yields  
\begin{equation}
R_{\mu\nu}-\frac{1}{2}Rg_{\mu\nu}=\frac{\kappa 
T_{\mu\nu}}{L'(\phi)}-V(\phi)g_{\mu\nu},
\label{EOF1}
\end{equation}
and variation with respect to $\phi$ gives
\begin{equation}
-2V'(\phi)=\kappa T\frac{L''(\phi)}{[L'(\phi)]^2},
\label{EOF2}
\end{equation}
where $T=T_{\mu\nu}g^{\mu\nu}$. 
Equations~(\ref{EOF1}) and~(\ref{EOF2}) are the equations of the
gravitational field in the Einstein frame~\cite{Pal3,OK}.
One can easily check that equation~(\ref{EOF2}) is not independent, it 
results from equations~(\ref{resc}), (\ref{pot}), and~(\ref{EOF1}).
From equation~(\ref{EOF1}) it follows that the tensor $T_{\mu\nu}$ is 
not covariantly conserved, unlike that in the
Jordan frame~\cite{Koi}.\footnote[1]{
By covariant conservation of a tensor, we mean that the
divergence of this tensor vanishes
using $\Gamma^{\,\,\rho}_{\mu\,\nu}$ in the
definition of the covariant derivative.
If instead of $\Gamma^{\,\,\rho}_{\mu\,\nu}$, we used  
Christoffel symbols associated with $\tilde{g}_{\mu\nu}$ 
in this definition, then the
energy-momentum tensor from equation~(\ref{EMT1}), which is associated with
$\tilde{g}_{\mu\nu}$ as well, would be automatically
conserved~\cite{LL,Koi}. 

Instead of $T_{\mu\nu}$, one might have interpreted the energy-momentum 
tensor generated by the geodesic metric $g_{\mu\nu}$ as the true 
energy-momentum tensor of matter. 
Such a tensor, equal to the first term on the right-hand side of 
equation~(\ref{EOF1}), is covariantly conserved according to the above 
definition, since the connection
$\Gamma^{\,\,\rho}_{\mu\,\nu}$ is compatible with $g_{\mu\nu}$. 
Yet this equation would imply the constancy of $V(\phi)$, which, with 
equation~(\ref{pot}), yields $L(\phi)=\phi$. Therefore, we would arrive at
general relativity with a cosmological constant, which is not interesting 
from a modified gravity perspective~\cite{OK}.}

Contracting equation~(\ref{EOF1}) with the metric tensor $g^{\mu\nu}$ gives
\begin{equation}
R=-\frac{\kappa T}{L'(\phi)}+4V(\phi),
\label{struc1}
\end{equation} 
which is an equation for $R$ since both $\phi$ and $T$ depend only on 
$R$ due to equations~(\ref{resc}) and~(\ref{EOF2}). 
This is equivalent to
\begin{equation}
\phi L'(\phi)-2L(\phi)=\kappa TL'(\phi),
\label{struc3}
\end{equation}
which is an equation for $\phi$ as a function of $T$, and called
the structural equation~\cite{Pal8}.
For the case of $T=0$ which holds at the early stages of the universe 
(relativistic matter) and, to good approximation, during advanced 
cosmic acceleration (when $\kappa T\ll |\phi|$), we obtain
\begin{equation}
\phi L'(\phi)-2L(\phi)=0,
\label{struc2}
\end{equation}
which agrees with the structural equation in vacuum for an $f(R)$ 
gravity in the 
Jordan frame~\cite{Pal2}.
Equation~(\ref{struc2}) gives $\phi=\,$const, which, upon substitution into 
equation~(\ref{EOF1}), leads to the Einstein equations of general relativity 
with a cosmological constant~\cite{univ} and the gravitational 
coupling $\kappa$ modified by a constant factor $L'(\phi)$.
Therefore, inflation and current cosmic acceleration can be generated by a 
metric-affine $f(R)$ gravity formulated in the Einstein frame 
if the gravitational Lagrangian $L(R)$ contains both positive and 
negative powers of the curvature scalar, because such a possibility
results from the structural equation.

Let us consider the Lagrangian~\cite{Pal7}, 
\begin{equation}
L(\phi)=\phi+\frac{\phi^3}{\beta^2}-\frac{\alpha^2}{3\phi},
\label{infacc1}
\end{equation}
where $\beta$ and $\alpha$ are positive constants, remembering that 
$\phi$ is the curvature scalar in the Jordan frame, $\tilde{R}$.
Equation~(\ref{struc2}) reads
\begin{equation}
\phi^4-\beta^2\phi^2+\alpha^2\beta^2=0,
\label{infacc2}
\end{equation} 
and for $\alpha\ll\beta$ it has two de Sitter solutions:
\begin{equation}
\phi_{inf}=-\beta,\,\,\phi_{ca}=-\alpha,
\label{infacc3}
\end{equation}
describing inflation and present acceleration, 
respectively~\cite{fR2,Pal7}.
The corresponding values of the curvature scalar in the Einstein frame 
are obtained with equation~(\ref{resc}):
\begin{equation}
R_{inf}=-\frac{\beta}{4},\,\,R_{ca}=-\frac{3\alpha}{4}.
\label{infacc4}
\end{equation}
If $T\neq0$, equation~(\ref{struc3}) leads to a quintic equation for $\phi$ 
as a function of $T$.

\section{Inflation and current cosmic acceleration in FLRW cosmology}
We now proceed to study a Friedmann-Lema\^{i}tre-Robertson-Walker 
(FLRW) cosmology driven by the above field equations. 
Let us consider a homogeneous and isotropic universe which is spatially 
flat~\cite{CMB2}.
In this case, the interval is given by
\begin{equation}
ds^2=c^2dt^2-a^2(t)(dx^2+dy^2+dz^2),
\label{FLRW}
\end{equation}
where $a(t)$ is the scale factor.
Moreover, the energy-momentum tensor of matter in the comoving 
frame of reference is of the form
\begin{equation}
T_{\mu}^{\nu}=diag(\epsilon,\,-P,\,-P,\,-P),
\label{EMT2}
\end{equation} 
where $\epsilon$ is the energy density and $P$ denotes the pressure.
Clearly, $T=\epsilon-3P$. In the matter-dominated universe
we can approximate matter by dust and set $P=0$.

The Hubble parameter $H=\frac{\dot{a}}{a}$ can be obtained from 
the $00$ component of
equation~(\ref{EOF1}), which for the metric~(\ref{FLRW}) and for dust becomes
\begin{equation}
H(\phi)=c\sqrt{\frac{\phi L'(\phi)-3L(\phi)}{6[L'(\phi)]^2}}.
\label{Hub1}
\end{equation}
The dot denotes the time-derivative and we used equation~(\ref{struc3}).
At very late stages of the universe expansion ($T\sim0$) 
equations~(\ref{struc3}) and~(\ref{Hub1}) give
\begin{equation}
H(\phi)=c\sqrt{-\frac{L(\phi)}{6[L'(\phi)]^2}}
\label{Fri1}
\end{equation}
For early stages we must use the ultrarelativistic equation
of state, $P=\frac{\epsilon}{3}$, for which $T$ vanishes as well.
In this case, the Hubble parameter is given by
\begin{equation}
H(\phi)=c\sqrt{\frac{L(\phi)-\phi L'(\phi)}{6[L'(\phi)]^2}}.
\label{Hub2}
\end{equation} 
If we use the Lagrangian~(\ref{infacc1}), the Einstein frame 
values of $H$ for the two de Sitter phases become
\begin{equation}
H_{inf}=c\sqrt{\beta/48},\,\,H_{ca}=c\sqrt{\alpha/16}.
\label{Fri2}
\end{equation}

For lower temperatures, matter becomes nonrelativistic and $T$ increases.
Since $T$ deviates from zero, $\phi$ ceases to be constant and the field 
equations differ from those with a cosmological constant.
The universe gradually becomes matter-dominated, and undergoes a transition 
from inflationary acceleration to a decelerated expansion phase.
At some moment, $T$ reaches a maximum and then decreases.
When $\kappa T$ becomes much smaller than $|\phi|$, the universe
undergoes a smooth transition back to an exponential acceleration.

We do not establish the kind of matter considered here since, for cosmological
purposes, matter in a homogeneous and isotropic universe can be simply 
described by the energy density and pressure related to each other by the 
effective quation of state. 
In the late universe, the kind of matter does not influence the 
deceleration-acceleration transition because matter is nonrelativistic ($P=0$).
For the early universe, the composition of matter is crucial to establish
when and how the transition from inflation to the matter-dominated epoch
occurs.
Since the purpose of this paper is to show that inflation and present cosmic 
acceleration {\it can} arise from nonlinear gravity in the Einstein frame, we
did not address the particle physics of the inflation-deceleration
transition.

For dust, the covariant conservation of the right-hand side 
of equation~(\ref{EOF1}) leads to
\begin{equation}
L'(\phi)\frac{d}{dt}\biggl[\frac{\epsilon(\phi)}{L'(\phi)}-\frac{V(\phi)}{\kappa}\biggr]+3H(\phi)\epsilon(\phi)=0,
\label{cons}
\end{equation}
which with equations~(\ref{pot}) and~(\ref{struc3}) reads
\begin{equation}
\dot{\phi}=\frac{6L'H(\phi L'-2L)}{2L'^2+\phi L'L''-6LL''}.
\label{phidot1}
\end{equation}
From now on, the prime denotes the $\phi$-derivative.
Therefore, making use of equation~(\ref{Hub1}) we obtain
\begin{equation}
\dot{\phi}=\frac{\sqrt{6}c(\phi L'-2L)\sqrt{\phi L'-3L}}{2L'^2+\phi 
L'L''-6LL''},
\label{phidot2}
\end{equation}
which can be integrated for a given function $L=L(\phi)$, yielding 
$\phi=\phi(t)$ and $R=R(t)$.
Combining equations~(\ref{Hub1}) and~(\ref{phidot2}) gives the function 
$H=H(t)$, from which we obtain the final expression for expansion, $a=a(t)$.

For the Lagrangian~(\ref{infacc1}), equation~(\ref{Hub1}) reads 
\begin{equation}
\frac{H^2}{c^2}=\frac{-\frac{\phi}{3}+\frac{2\alpha^2}{9\phi}}{(1+\frac{3\phi^2}{\beta^2}+\frac{\alpha^2}{3\phi^2})^2},
\label{Hub3}
\end{equation}
and equation~(\ref{phidot2}) becomes
\begin{equation}
\dot{\phi}=\frac{c(-\phi+\frac{\phi^3}{\beta^2}+\frac{\alpha^2}{\phi})\sqrt{\frac{2\alpha^2}{\phi}-3\phi}}{1-\frac{2\alpha^4}{3\phi^4}-\frac{9\phi^2}{\beta^2}+\frac{7\alpha^2}{3\phi^2}+\frac{10\alpha^2}{\beta^2}}.
\label{phidot3}
\end{equation}
In the Jordan frame, the equation for $\dot{\phi}$ is more complicated 
(see~\cite{Pal7}). 
From the time-dependence of $\phi$ we can derive the time-dependence of 
$R$ by simply applying equation~(\ref{resc}).
We give the approximate expressions for $\dot{\phi}$ for three regions: 
inflation ($\phi\sim-\beta$), the matter-dominated era 
($-\alpha\gg\phi\gg-\beta$), and advanced cosmic acceleration 
($\phi\sim-\alpha$). 
In the course of expansion, the quantity $\phi$ varies between $-\beta$ 
and $-\alpha$ (the two de Sitter values).

In the first case, we use $\phi\ll-\alpha$ to obtain
\begin{equation}
\frac{H^2}{c^2}=-\frac{\phi}{3(1+\frac{3\phi^2}{\beta^2})^2}
\label{Hinf}
\end{equation}
and
\begin{equation}
\dot{\phi}=\frac{c(-\phi+\frac{\phi^3}{\beta^2})\sqrt{-3\phi}}{1-\frac{9\phi^2}{\beta^2}}.
\label{pdinf}
\end{equation}
We see that this expression becomes singular at 
$\phi=-\frac{\beta}{3}$. 
We note, however, that in our derivation of equation~(\ref{phidot2}) we 
assumed $P=0$. 
Therefore, the condition $\phi=-\frac{\beta}{3}$ is the limit of 
validity of this assumption, i.e., may be regarded as the limit of 
nonrelativisticity of matter. 
In order to derive the universe expansion for smaller values of $\phi$ 
(down to $-\beta$), we must use a relativistic equation of state, for 
which $P\neq0$.

In the second case, we neglect all terms with $\alpha$ or $\beta$, 
arriving at
\begin{equation}
\frac{H^2}{c^2}=-\frac{\phi}{3},\,\,\,\dot{\phi}=\sqrt{3}c(-\phi)^{3/2}.
\label{Hmat}
\end{equation}
Since now $L'(\phi)\approx 1$ and $\phi\approx R$, the above formulae 
reproduce the standard Friedmann cosmology:
$R\propto\epsilon\propto t^{-2},\,\,a\propto t^{2/3}$.
Therefore, this region of time also corresponds to the 
radiation-dominated era and big bang nucleosynthesis. 
Note that the Friedmann equations may be obtained as a limiting case 
only if $\alpha\ll\beta$.

In the third case, we have
\begin{equation}
\frac{H^2}{c^2}=\frac{-\frac{\phi}{3}+\frac{2\alpha^2}{9\phi}}{(1+\frac{\alpha^2}{3\phi^2})^2},
\label{Hca}
\end{equation}
and equation~(\ref{phidot2}) gives
\begin{equation}
\dot{\phi}=\frac{c(-\phi+\frac{\alpha^2}{\phi})\sqrt{\frac{2\alpha^2}{\phi}-3\phi}}{1-\frac{2\alpha^4}{3\phi^4}+\frac{7\alpha^2}{3\phi^2}}.
\label{phidotca}
\end{equation}
In the course of time $\phi\rightarrow-\alpha$ and 
$\dot{\phi}\rightarrow0$, and the universe asymptotically approaches a 
de Sitter 
expansion~\cite{Pal2}.

Finally, we derive the Einstein equations in the universe at very late
stages of its expansion (when $\kappa T\ll\alpha$) in the linear and quadratic
approximation of a small quantity $\frac{\kappa T}{\alpha}$.
Equation~(\ref{struc3}) becomes cubic in $\phi$,
\begin{equation}
\phi^3+\kappa T\phi^2-\alpha^2\phi+\frac{\alpha^2\kappa T}{3},
\label{struc4}
\end{equation}
and its solution, which deviates from $-\alpha$ by a term linear
in $\frac{\kappa T}{\alpha}$, is
\begin{equation}
\phi=-\alpha-\frac{2}{3}\kappa T+O(T^2).
\label{struc5}
\end{equation}
In this approximation, we obtain
\begin{equation}
L(\phi)=-\frac{2\alpha}{3}\Bigl(1+\frac{4\kappa T}{3\alpha}\Bigr),\,\,\,L'(\phi)=\frac{4}{3}\Bigl(1-\frac{\kappa T}{6\alpha}\Bigr),
\label{struc6}
\end{equation}
and the Einstein equations become
\begin{equation}
R_{\mu\nu}-\frac{1}{2}Rg_{\mu\nu}=\frac{3}{4}\kappa T_{\mu\nu}+\Lambda g_{\mu\nu},
\label{EOF3}
\end{equation}
where 
\begin{equation}
\Lambda=\frac{3\alpha}{16}
\end{equation}
plays the role of a cosmological constant.
We see that the coupling between matter and the gravitational field 
is decreased by a factor of $3/4$, as in the Jordan frame~\cite{Pal6}.
The conservation law for the energy density~(\ref{cons}) in this limit
becomes
\begin{equation}
\dot{\epsilon}+3H\epsilon=0,
\label{cons2}
\end{equation}   
which gives the usual scaling of nonrelativistic matter, 
$\epsilon\propto a^{-3}$, as in the $\Lambda CDM$ model.

The solution of equation~(\ref{struc3}), which deviates from $-\alpha$ 
by terms linear and quadratic in $\frac{\kappa T}{\alpha}$, appears to be the same as that in the linear approximation:
\begin{equation}
\phi=-\alpha-\frac{2}{3}\kappa T+O(T^3).
\label{struc7}
\end{equation}
Consequently, we obtain
\begin{eqnarray}
& & L'(\phi)=\frac{4}{3}\Bigl(1-\frac{\kappa T}{3\alpha}+\frac{\kappa^2 T^2}{3\alpha^2}\Bigr), \nonumber \\
& & V(\phi)=-\frac{3\alpha}{16}+\frac{\kappa^2 T^2}{16\alpha},
\label{struc8}
\end{eqnarray}
and the Einstein equations become
\begin{equation}
R_{\mu\nu}-\frac{1}{2}Rg_{\mu\nu}=\frac{3}{4}\kappa T_{\mu\nu}\Bigl(1+\frac{\kappa T}{3\alpha}\Bigr)+\Lambda g_{\mu\nu},
\label{EOF4}
\end{equation}
where the cosmological constant is now a function of $T$,
\begin{equation}
\Lambda=\frac{3\alpha}{16}-\frac{\kappa^2 T^2}{16\alpha}.
\end{equation}
The coupling between matter and the gravtitational field 
is decreased by a factor which tends to $3/4$ as $T\rightarrow0$.
The conservation law for the energy density~(\ref{cons}) becomes
\begin{equation}
\frac{d}{dt}\Bigl(\epsilon+\frac{\kappa \epsilon^2}{4\alpha}\Bigr)+H\Bigl(3\epsilon+\frac{\kappa \epsilon^2}{\alpha}\Bigr)=0,
\label{cons3}
\end{equation}   
which gives the energy density scaling that differs from the $\Lambda CDM$ scaling, and tends to it as $\epsilon\rightarrow0$ ($\kappa\epsilon\ll\alpha$).

\section{Summary}
Inflation and current cosmic acceleration can be generated by replacing the GR 
Einstein-Hilbert Lagrangian with a modified $f(R)$ Lagrangian, which 
offers an alternative explanation of cosmological acceleration.
We used the metric-affine variational formalism, and chose the Einstein 
conformal frame as being physical.
In this frame, we derived explicit formulae for the 
matter-dominated universe expansion for the particular case 
$L(R)=R+\frac{R^3}{\beta^2}-\frac{\alpha^2}{3R}$, and showed that they 
reproduce the standard 
Friedmann cosmology in the region of middle values of $R$. 
We also demonstrated that gravitational effects of matter in such a universe 
at very late stages of its expansion
are weakened by a factor that tends to the Jordan frame value $3/4$.
Finally, we showed that the energy density of matter in the late universe 
scales according to the same power law as in the $\Lambda CDM$ model
only in the limit $\epsilon\rightarrow0$.

We did not address the physical scenario of the inflation-deceleration
transition.
We also did not address the problem of constraining the values of the 
constants $\alpha$ and $\beta$ from astronomical observations. 
Such studies, as well as determination of the function $f(R)$ from the data, 
as has been explored in the metric approach~\cite{obs}, will be the subject 
of future work.

\end{document}